\documentclass[12pt]{article}
\usepackage{epsfig}

\newcommand\pubnumber{}
\newcommand\pubdate{December, 2000}
\newcommand\hepnumber{hep-ph/0012316}

\def\csumb{Department of Physics and Astronomy\\
Arizona State University, Tempe, AZ 85287 USA}
\def\support{\footnote{Work supported by the Department of Energy
under Grant No.\ DE-AC05-84ER40150.}} 

\def\Title#1{\begin{center} {\Large\bf #1 } \end{center}}
\def\Author#1{\begin{center}{ \sc #1} \end{center}}
\def\Address#1{\begin{center}{ \it #1} \end{center}}

\newcommand\pubblock{\rightline{\begin{tabular}{l} \pubnumber\\
         \pubdate\\ \hepnumber \end{tabular}}}
\newenvironment{Abstract}{\begin{quotation}  }{\end{quotation}}
\newenvironment{Presented}{\begin{quotation} \begin{center} 
             Presented at the\end{center}
      \begin{center}\begin{large}}{\end{large}\end{center} \end{quotation}}
\def\Acknowledgments{\bigskip  \bigskip \begin{center}
          \large\bf Acknowledgments\end{center}}

\makeatletter
\def\section{\@startsection{section}{0}{\z@}{5.5ex plus .5ex minus
 1.5ex}{2.3ex plus .2ex}{\large\bf}}
\def\subsection{\@startsection{subsection}{1}{\z@}{3.5ex plus .5ex minus
 1.5ex}{1.3ex plus .2ex}{\normalsize\bf}}
\def\subsubsection{\@startsection{subsubsection}{2}{\z@}{-3.5ex plus
-1ex minus  -.2ex}{2.3ex plus .2ex}{\normalsize\sl}}

\renewcommand{\@makecaption}[2]{%
   \vskip 10pt
   \setbox\@tempboxa\hbox{\small #1: #2}
   \ifdim \wd\@tempboxa >\hsize     
       \small #1: #2\par          
     \else                        
       \hbox to\hsize{\hfil\box\@tempboxa\hfil}
   \fi}

 \def\citenum#1{{\def\@cite##1##2{##1}\cite{#1}}}
 
\newcount\@tempcntc
\def\@citex[#1]#2{\if@filesw\immediate\write\@auxout{\string\citation{#2}}\fi
  \@tempcnta\z@\@tempcntb\m@ne\def\@citea{}\@cite{\@for\@citeb:=#2\do
    {\@ifundefined
       {b@\@citeb}{\@citeo\@tempcntb\m@ne\@citea\def\@citea{,}{\bf ?}\@warning
       {Citation `\@citeb' on page \thepage \space undefined}}%
    {\setbox\z@\hbox{\global\@tempcntc0\csname b@\@citeb\endcsname\relax}%
     \ifnum\@tempcntc=\z@ \@citeo\@tempcntb\m@ne
       \@citea\def\@citea{,}\hbox{\csname b@\@citeb\endcsname}%
     \else
      \advance\@tempcntb\@ne
      \ifnum\@tempcntb=\@tempcntc
      \else\advance\@tempcntb\m@ne\@citeo
      \@tempcnta\@tempcntc\@tempcntb\@tempcntc\fi\fi}}\@citeo}{#1}}
\def\@citeo{\ifnum\@tempcnta>\@tempcntb\else\@citea\def\@citea{,}%
  \ifnum\@tempcnta=\@tempcntb\the\@tempcnta\else
  {\advance\@tempcnta\@ne\ifnum\@tempcnta=\@tempcntb \else\def\@citea{--}\fi
    \advance\@tempcnta\m@ne\the\@tempcnta\@citea\the\@tempcntb}\fi\fi}
\makeatother

\def\beq{\begin{equation}}
\def\eeq#1{\label{#1}\end{equation}}
\def\eeqn{\end{equation}}

\newenvironment{Eqnarray}%
   {\arraycolsep 0.14em\begin{eqnarray}}{\end{eqnarray}}
\def\beqa{\begin{Eqnarray}}
\def\eeqa#1{\label{#1}\end{Eqnarray}}
\def\eeqan{\end{Eqnarray}}

\let\bar=\overbar

\def\Dslash{\not{\hbox{\kern-4pt $D$}}}
\def\dslash{\not{\hbox{\kern-2pt $\del$}}}

\def\msb{{\bar{\ssstyle M \kern -1pt S}}}

\def\lsim{\mathrel{\raise.3ex\hbox{$<$\kern-.75em\lower1ex\hbox{$\sim$}}}}
\def\gsim{\mathrel{\raise.3ex\hbox{$>$\kern-.75em\lower1ex\hbox{$\sim$}}}}

\begin{document}
\begin{titlepage}
\pubblock

\vfill
\def\thefootnote{\fnsymbol{footnote}}
\Title{On Radiative Weak Annihilation Decays}
\vfill
\Author{Richard F. Lebed\support}
\Address{\csumb}
\vfill
\begin{Abstract}
We discuss a little-studied class of weak decay modes sensitive to
only one quark topology at leading order in $G_F$: $M \to m
\gamma$, where $M,m$ are mesons with completely distinct flavor quantum
numbers.  Specifically, they proceed via the annihilation of the
valence quarks through a $W$ and the emission of a single hard photon,
and thus provide a clear separation between CKM and strong interaction
physics. We survey relevant calculations performed to date, discuss
experimental discovery potential, and indicate interesting future
directions.
\end{Abstract}
\vfill
\begin{Presented}
5th International Symposium on Radiative Corrections \\ 
(RADCOR--2000) \\[4pt]
Carmel CA, USA, 11--15 September, 2000
\end{Presented}
\vfill
\end{titlepage}
\def\thefootnote{\arabic{footnote}}
\setcounter{footnote}{0}

The feature that makes heavy quark physics appealing---the decoupling
of the heavy quark matrix elements from those of the light degrees of
freedom---can prove to be treacherous if one cannot track the
processes through which the quarks of various flavors are created or
destroyed.  Such ambiguities plague the extraction of
Cabibbo-Kobayashi-Maskawa (CKM) elements from nonleptonic weak decays.

In this light, processes with unusual flavor quantum numbers are
useful since they serve to distinguish the weak interaction and strong
interaction physics: As seen below, unusual flavor quantum numbers
imply a very limited number of possible Feynman diagram topologies.
The price one must pay for this clarity is that such interesting
decays tend to be quite rare.  In particular, the modes $M \to m
\gamma$ discussed in this talk are radiative (rates $\propto
\alpha_{\rm EM}$) weak ($\propto G_F^2 | V^* V |^2$) processes with
pointlike annihilation ($\propto f_M/M \cdot f_m/m$, where $f$
indicates the meson decay constant).  Nevertheless, we argue below
that once such modes are produced, they should be relatively easy to
detect.

\begin{figure}[b!]
\begin{center}
\epsfig{file=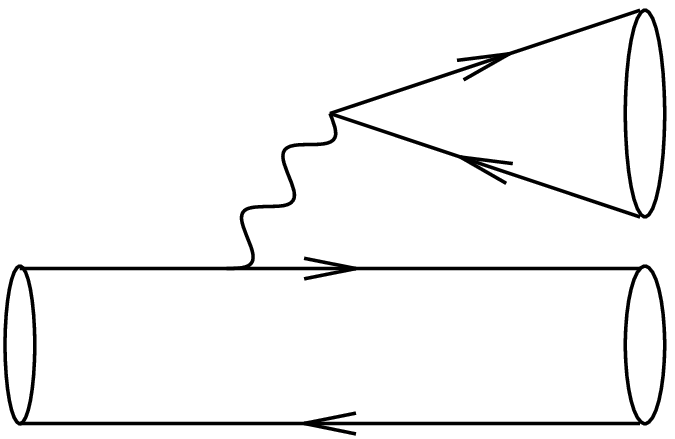,height=1.5in}\hfil\epsfig{file=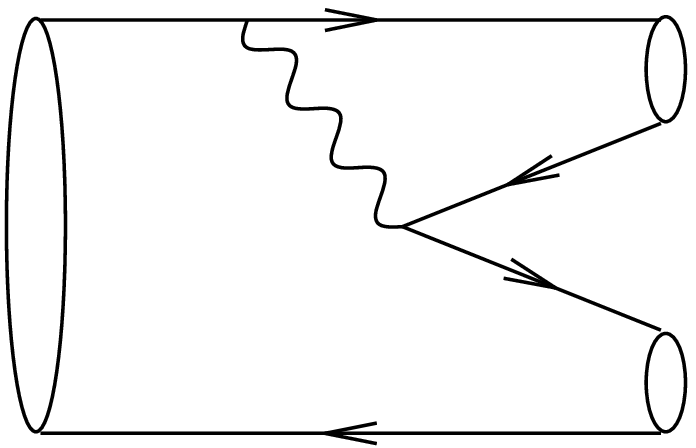,height=1.0in}
\vskip 0.2in
\epsfig{file=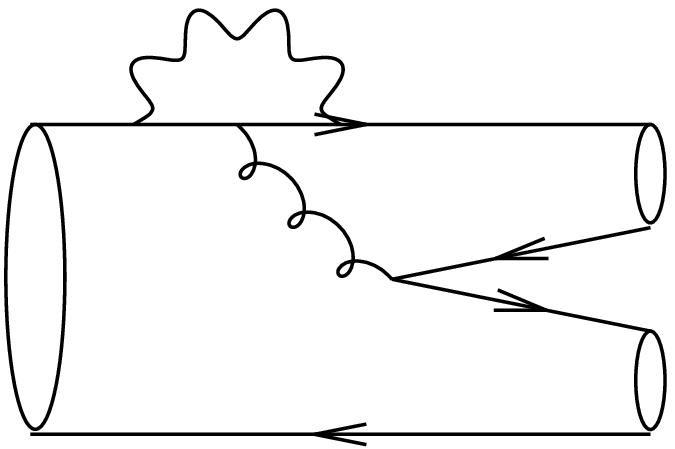,height=1.5in}\hfil\epsfig{file=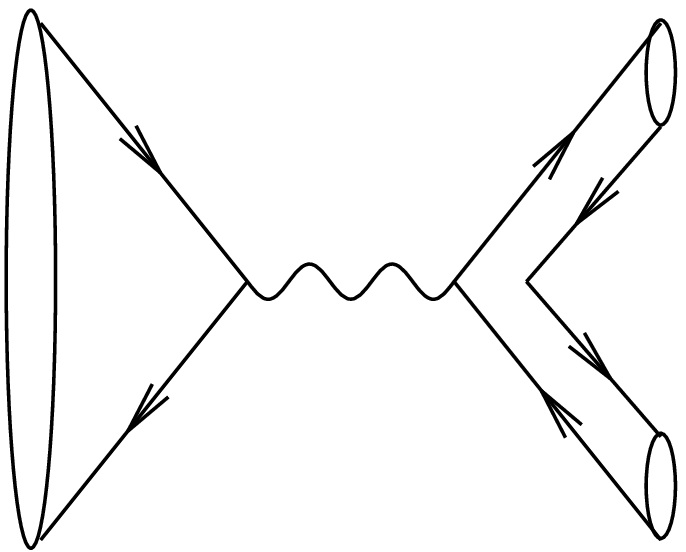,height=1.5in}
\vskip 0.2in
\epsfig{file=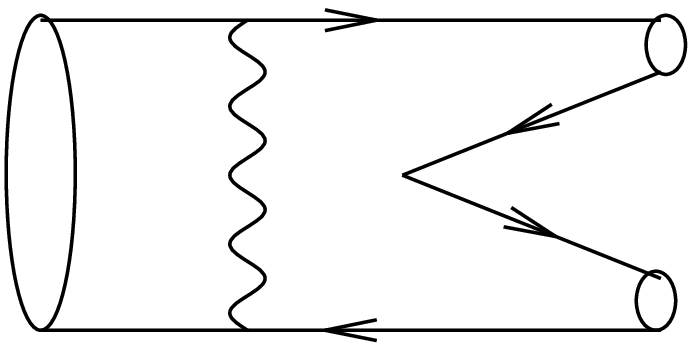,height=1.0in}\hfil\epsfig{file=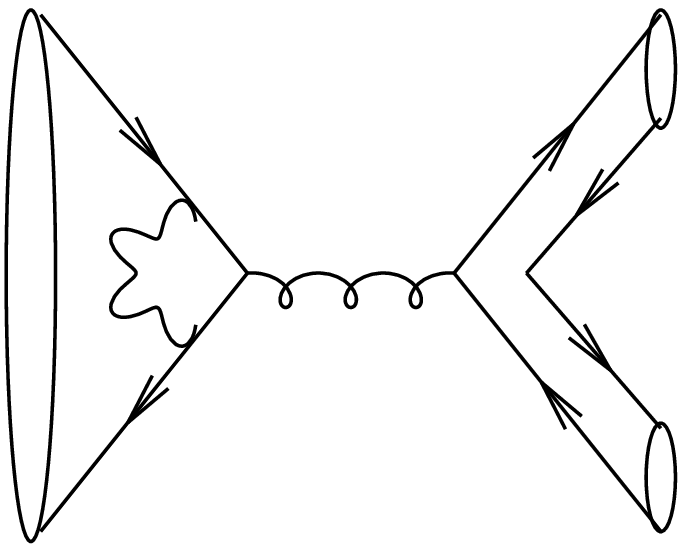,height=1.5in}

\caption[0]{The six flavor-changing weak decay topologies at
$O(G_F^1)$.  Reading across and from top to bottom, these are labeled
$T$, $C$, $P$, $A$, $E$, and $PA$, respectively.  Ovals indicate
hadronization into color-singlet mesons.  For processes with the $T$
topology, there is often also a $C$ diagram contribution, so that the
two classes can mix through final-state interactions.  The $P$, $A$,
$E$, and $PA$ diagrams have variants in which the the $q\bar q$ pair
from the vacuum hadronize into a single flavor-singlet meson.}
\label{fig:diags}
\end{center}
\end{figure}

To study the flow of flavor in a flavor-changing weak decay process,
it is sufficient to work at the partonic level with simple quark
diagrams, since gluons and sea quark pairs carry only flavor-singlet
quantum numbers.  Only valence quarks and vacuum-produced $q\bar q$
pairs that become valence quarks need be considered.  Thus, the
complications of QCD are irrelevant if one wishes only to classify
weak decay processes.  As shown in Ref.~\cite{GHLR}, only six such
classes exist at $O(G_F^1)$, since only the $W$ boson changes flavors
in the standard model.  These classes are $T$ (color-unsuppressed
tree), $C$ (color-suppressed tree), $P$ (penguin), $A$ (weak
annihilation), $E$ (weak exchange), and $PA$ (penguin annihilation)
diagrams, as depicted in Fig.~\ref{fig:diags}.

One may now enumerate a number of problems inherent to computing weak
meson decays of the form $M \to m_1 m_2$.  The first such difficulty
is the most obvious and endemic to any hadronic process, namely, that
hadron wavefunctions are not precisely known and must be modeled in
order for a calculation to be performed.  Second, for electrically
neutral mesons---even with valence quarks with distinct flavors---the
asymptotic states are not pure flavor eigenstates, and then one must
take into account $K\bar K$, $D \bar D$, $B \bar B$, or $B_s \bar B_s$
mixing.

In addition, however, there are complications best seen by using the
diagrammatic classification.  While it is true that any arbitrarily
complicated Feynman diagram for a flavor-changing $M \to m_1 m_2$
meson decay falls uniquely into one of these six classes, it is also
true that any such decay tends to have contributions from more than
one topology.  In particular, processes that have a $T$ diagram often
also have a $C$ diagram, and the two can mix under final-state
interactions (FSI's).  As an example, consider $B^+ \to \pi^+ \bar D^0
= \bar b u \to (u \bar d) (\bar c u)$.  The weak decay at the quark
level may proceed through $\bar b \to \bar c W^+ \to \bar c u \bar d$,
and the $u$ valence quark in $\pi^+$ may either come from the weak
vertex or the spectator.  In other cases, valence quarks may emerge
from pair creation due to fragmentation.  The basic problem here is
one of redundant quark flavor in the final state: Since two $u$ quarks
are indistinguishable, one is faced with the problem of where each one
originates, and this redundancy is forced by the limited number of
distinct quark flavors available for hadron formation.

Another affliction of $M \to m_1 m_2$ best seen in terms of the
diagram topologies is the problem of ``generalized penguin
pollution,'' which we define to be the contribution to a decay from at
least two diagrams containing different CKM couplings.  A classic
example is the decay $B^0 \to \pi^+ \pi^- = \bar b d \to (u \bar d) (d
\bar u)$.  In the $T$ diagram, the weak decay at the quark level is
$\bar b \to \bar u W^+ \to \bar u u \bar d$, where the $\bar u$ quark
hadronizes with the spectator $d$ quark, and the CKM coefficient is
$V_{ub}^* V_{ud}^{\vphantom{\dagger}}$.  On the other hand, a $P$
diagram may also contribute (hence the original name ``penguin
pollution''), in which $\bar b \to \bar d g \to \bar d u \bar u$.  The
penguin loop, dominated by the top quark contribution, produces
primarily the CKM coupling $V_{tb}^* V_{td}^{\vphantom{\dagger}}$.
Again, the ultimate reason that a large proportion of possible modes
exhibit generalized penguin pollution is the existence of a limited
number of quark flavors available for the decay: In the example, the
$u \bar u$ pair can either emerge from a strong or weak process.

Typically, $M \to m_1 m_2$ meson decays tend to suffer at least one of
the latter two problems.  It is difficult to find modes proceeding
through only one topology, chiefly owing to the limited number of
distinct quark flavors; in particular, $M \to m_1 m_2$ contains six
quarks, while only the lightest five quark flavors form mesons.

To evade this problem, let us consider instead the meson decays $M \to
m \gamma$~\cite{RFL}.  Here one has only four quarks, which can easily
be chosen distinct.  In this case there is clearly (as a few moments'
study of Fig.~\ref{fig:diags} should convince the reader) only one
weak topology available at $O(G_F^1)$: If the meson is charged, then
$M^+ \to m^+ \gamma$ proceeds uniquely through the weak annihilation
($A$) diagram, while if the meson is neutral, then $M^0 \to m^0
\gamma$ proceeds uniquely through the weak exchange ($E$) diagram.  To
date, no such decays have been observed, so an order-of-magnitude
calculation of their rates serves to guide not only future
calculations, but experimental searches as well.  The photon may be
attached to any charged particle line, and is hard and monochromatic,
fixed in energy due to the restrictive kinematics of two-body decays.
The $E$ processes are certainly interesting, but introduce the problem
of $M^0 \bar M^0$ and $m^0 \bar m^0$ mixing mentioned above, so we
concentrate below on the $A$ processes.

\begin{figure}[b!]
\begin{center}
\epsfig{file=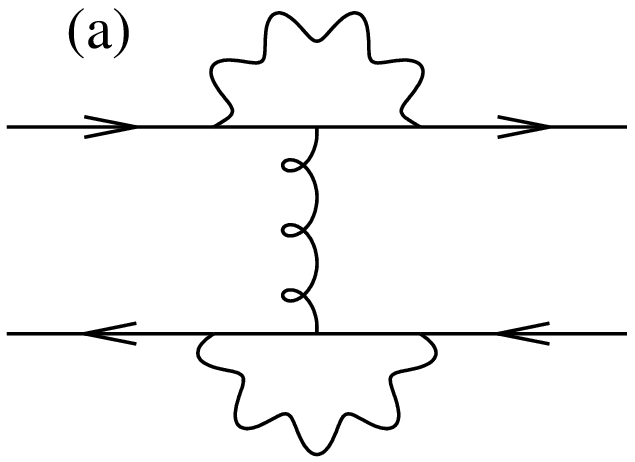,height=1.5in}\hfil\epsfig{file=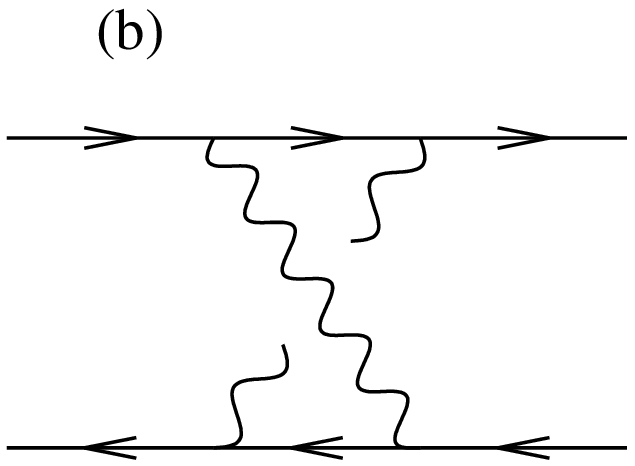,height=1.5in}
\caption[0]{Diagrams contributing at $O(G_F^2)$ to processes described
at $O(G_F^1)$ by the $A$ diagram of Fig.~\ref{fig:diags}.  These are
the ($a$) ``di-penguin'' and ($b$) crossed-box diagrams.  The gluons
emerging from the $W$ loops in the di-penguin need not be the same;
indeed, the two penguins can be separated by long-distance effects.}
\label{fig:gf2}
\end{center}
\end{figure}

Thus far we have considered contributions only at $O(G_F^1)$.  One may
also neglect the diagram in which the photon couples to the $W^{\pm}$,
since it produces an extra $1/M_W^2$ propagator suppression; thus, one
need consider only diagrams in which the photon couples to one of the
quarks.  At a similar numerical size for $A$ processes are the
$O(G_F^2)$ diagrams depicted in Fig.~\ref{fig:gf2}.  These consist of
``di-penguin'' and crossed-box diagrams; however, neither class is
expected to be particularly large, since the enhanced significance of
ordinary penguin and box diagrams occurs due to virtual $t$ quark
lines in the loops.  In the case of charged mesons, one of the valence
quarks is necessarily $u$-type, and the virtual quark connecting to it
through a $W$ vertex is necessarily $d$-type, and thus does not give
rise to a large contribution.  One concludes that, in the standard
model at least, the $O(G_F^1)$ contribution should dominate the rate
for $A$ processes.

This begs the question of whether any common non-standard physics
might contribute to, or even dominate, $A$ processes.  Two very simple
possibilities jump immediately to mind.  First, the $s$-channel
exchange of a $W$ may be replaced with the $t$-channel exchange of a
flavor-changing neutral current (FCNC) boson $X$.  Such a process is
important if
\begin{equation}
\frac{V_1 V_2}{M_W^2} \lsim \frac{g_1 g_2}{M_X^2},
\end{equation}
where $V_i$ and $g_i$ represent the CKM and new physics couplings,
respectively, at the two vertices.  The best potential for new physics
discovery is when $g_i$ are $O(1)$ and $V_i$ are as small as possible.
For example, in the case $B^+ \to D^{*+} \gamma$, $V_1 = V_{ub}^* \sim
\lambda^3$ and $V_2 = V_{cd} \approx \lambda$, with Wolfenstein
$\lambda \approx 0.2$.  Then one immediately finds
\begin{equation}
M_X \lsim 2 \ {\rm TeV},
\end{equation}
a fairly stringent bound, considering that one must still make sure
that $K \bar K$ mixing and other FCNC constraints are properly taken
into account.

Another possibility is the $s$-channel exchange of a charged Higgs
boson.  Since the current lower limit~\cite{PDG} is $M_{H^+} > 130$
GeV, these decays are a possible place to find new physics (using the
same tree-level estimation as above) if the corresponding $H^+ q
\bar q$ Yukawa couplings are not smaller than the CKM elements $V_1
V_2$.

Having discussed the restrictive nature of flavor distinctiveness on
generating processes with unique flavor topologies, it is natural to
present the complete list~\cite{RFL} of such decays for pure $A$
processes.  One must choose two from the list $\{b,s,d\}$ and two from
the list $\{c,u\}$, for a total of six possibilities.  Only the
lightest pseudoscalar $M$ of each flavor content decays dominantly
weakly.  Then angular momentum conservation requires that the spin of
the photon must be balanced by a daughter meson $m$ of spin $\ge 1$;
for sake of illustration, we take $m$ to be the lowest-lying vector
meson, which should presumably boast the largest transition rate for
any state with the given final flavor quantum numbers.
Table~\ref{modes} presents the modes just described, along with the
corresponding CKM coefficients.

\def\tableline{\noalign{
\hrule height.7pt depth0pt\vskip3pt}}

\begin{table}[t!]
\caption{\label{modes}Flavor structure and mesonic decay modes of weak
annihilation radiative decays.  The CKM coefficient for each
process is accompanied by its magnitude in powers of Wolfenstein
$\lambda \approx 0.2$.}

\vskip 2ex

\begin{center}
\setlength{\tabcolsep}{9pt}
\renewcommand{\arraystretch}{1.2}
\begin{tabular}{lcccc}
\tableline
Valence structure & Decay mode && CKM Elements & \\
\hline

$\bar b u \to c \bar s \gamma$ & $B^+ \to D_s^{*+} \gamma$ &&
$V^*_{ub} V^{\vphantom{\dagger}}_{cs} \sim \lambda^3$ & \\

$\bar b u \to c \bar d \gamma$ & $B^+ \to D^{*+} \gamma$ && $V^*_{ub}
V^{\vphantom{\dagger}}_{cd} \sim \lambda^4$ & \\

$\bar b c \to u \bar s \gamma$ & $B_c^+ \to K^{*+} \gamma$ && $V^*_{cb}
V^{\vphantom{\dagger}}_{us} \sim \lambda^3$ & \\

$\bar b c \to d \bar u \gamma$ & $B_c^+ \to \rho^+ \gamma$ && $V^*_{cb}
V^{\vphantom{\dagger}}_{ud} \sim \lambda^2$ & \\

$c \bar d \to u \bar s \gamma$ & $D^+ \to K^{*+} \gamma$ && $V^*_{cd}
V^{\vphantom{\dagger}}_{us} \sim \lambda^2$ & \\

$c \bar s \to u \bar d \gamma$ & $D_s^+ \to \rho^+ \gamma$ && $V^*_{cs}
V^{\vphantom{\dagger}}_{ud} \sim \lambda^0$ & \\
\hline

\end{tabular}
\end{center}
\end{table}


Note that the CKM suppression of the decays varies widely, from none
in the case of $D_s^+ \to \rho^+ \gamma$ to $\lambda^4$ in the case of
$B^+ \to D^{*+} \gamma$.  These decays appear to have been discussed
only rarely in the literature.  $D_s^+ \to \rho^+ \gamma$ has been
considered using the quark model~\cite{AK}, pole and vector meson
dominance methods~\cite{BGHP}, light-cone techniques~\cite{KSW}, and
effective field theory~\cite{Fajfer}.  The double Cabibbo-suppressed
mode $D^+ \to K^{*+} \gamma$, interesting since it is a neutrinoless
decay sensitive to $|V_{cs}|$, was also considered in
Refs.~\cite{BGHP,Fajfer}.  The modes $B^+ \to D_s^{*+} \gamma$ and
$D^{*+} \gamma$ (collectively, $D^{*+}_{(s)} \gamma$) were first
considered in Ref.~\cite{GL}, where they were suggested as possible
probes of $|V_{ub}|$.  The modes $B_c \to \rho^+ \gamma$ and $K^{*+}
\gamma$ were first considered~\cite{AS} in the context of light-cone
sum rules.

Let us consider in further detail the calculation of Ref.~\cite{GL}
since its methods and approximations figure large in the rest of this
talk.  In \cite{GL}, heavy quark effective theory (HQET) and
light-quark SU(3) are used to relate the four-fermion vertex $(\bar b
u)(c \bar d)$ or $(\bar b u)(c \bar s)$ appearing in the decays $B^+
\to D^{*+}_{(s)} \gamma$ to the vertex $(\bar b d)(b \bar d)$ that
appears in $B\bar B$ mixing.  Such an approach of course neglects the
``bag parameter'' $B$ ({\it i.e.}, the multiplicative long-distance
correction to factorization) relevant to each vertex, as well as the
mixing and short-distance renormalization of the two different color
Fierz orderings of the four-fermion operator.

\begin{figure}[b!]
\begin{center}
\epsfig{file=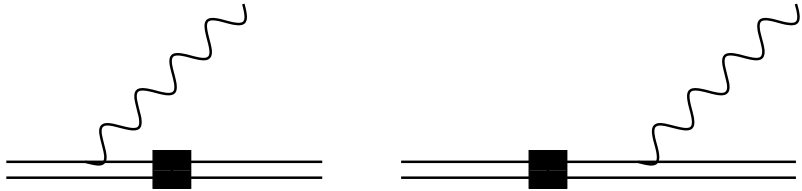,height=1.0in}
\caption[0]{Diagrams for $B^+ \to D^{*+}_{(s)}$, assuming dominance of
long-distance physics by single-meson states.  The square indicates
the weak interaction vertex.}
\label{fig:mesons}
\end{center}
\end{figure}

The next problem is how to incorporate long-distance effects between
the weak and electromagnetic vertices.  Here, the simplest ansatz is
adopted: One assumes that only the lightest meson propagates between
the two vertices with the same flavor quantum numbers as the meson on
the other side of the photon vertex, and the same spin-parity as the
meson on the other side of the weak vertex, .  This is depicted in
Fig.~\ref{fig:mesons}.  In each diagram the external states are $B^+$
and $D_{(s)}^{*+}$, while in the first the intermediate meson is
$B^{*+}$, and in the second it is $D^+_{(s)}$.  This approximation not
only neglects all higher resonances that may contribute in the
intermediate state (for example, $D_{(s)} (2S)$), but also
vector-dominance diagrams in which the photon is generated by a
resonance of a valence quark from one of the mesons and an antiquark
from the weak vertex (such as $B^+ \to D_s^{*+} \rho^0 \to D_s^{*+}
\gamma$), and multiparticle intermediates (such as $B^+ \to D^0 K^+
\to D^{*+}_s \gamma$) in which FSI's play an important role.

Finally, HQET is used to relate the $BB^{*}\gamma$ and
$D_{(s)}D^*_{(s)}\gamma$ couplings to $\Gamma (D^{*+} \to D^+
\gamma)$, for which an experimental upper bound exists~\cite{PDG}.
One finds the branching ratios (BR's)
\begin{eqnarray}
\lefteqn{{\rm BR}(B^+ \to D_s^{*+} \gamma)} && \nonumber \\
& = & 2 \times 10^{-7} \left(
\frac{B_B}{0.98} \right)^2 \left| \frac{V^*_{ub} V_{cs}}{3 \times
10^{-3}} \right| \left( \frac{\Gamma (D^{*+})}{0.131 \, {\rm MeV}}
\right) \left( \frac{{\rm BR} (D^{*+} \to D^+ \gamma)}{3.2\%} \right)
, \nonumber \\
\lefteqn{{\rm BR}(B^+ \to D^{*+} \gamma)} && \nonumber \\
& = & 7 \times 10^{-9} \left(
\frac{B_B}{0.98} \right)^2 \left| \frac{V^*_{ub} V_{cd}}{6.6 \times
10^{-4}} \right| \left( \frac{\Gamma (D^{*+})}{0.131 \, {\rm MeV}}
\right) \left( \frac{{\rm BR} (D^{*+} \to D^+ \gamma)}{3.2\%} \right)
. \nonumber \\ & &
\end{eqnarray}

The final approximation, using an on-shell electromagnetic coupling
(the $D^{*+} \to D^+ \gamma$ transition magnetic moment) to extract
the coupling of an intermediate meson, off-shell by the large amount
$m^2 (B^+) - m^2 (D_{(s)}^+)$ in the second diagram (as fixed by the
fact that the real photon has $q^2=0$), deserves special comment.  In
the original calculation~\cite{GL}, a new formal heavy quark limit
$m_b - m_c \lsim \Lambda_{\rm QCD} \ll m_{c,b}$ was invented, for
which the virtuality of the electromagnetic coupling is parametrically
small.  It is of course possible to remove this assumption when the
photon is virtual, as in $B^+ \to D_{(s)}^{*+} e^+ e^-$; then one may
find a kinematic region where the intermediate meson is approximately
at rest~\cite{EGN1}, or do even better and develop an operator product
expansion (OPE) in the variable $q^2 \gg \Lambda^2_{\rm QCD}$.  Then
one calculates~\cite{EGN2}, for example,
\begin{equation}
\left. {\rm BR} (B^+ \to D_s^{*+} e^+ e^-) \right|_{q^2 > 1 \, {\rm
GeV^2}} \approx 1.8 \times 10^{-9} .
\end{equation}
The authors of Refs.~\cite{EGN1,EGN2} also tackle the problem of
radiative weak exchange (the $E$ rather than $A$ diagram) exclusive
processes using the OPE approach in Ref.~\cite{EGN3}, finding similar
branching ratios.  The much smaller rates for $e^+e^-$ processes
compared to those for on-shell photon processes of course arise from
the additional factor of $\alpha_{\rm EM}$ from conversion of the
virtual photon.

One may also consider a simultaneous calculation~\cite{RFL} of all of
the decays in Table~\ref{modes} by using an approach similar to that
of Ref.~\cite{GL} but dropping the heavy quark approximations.  In
this case, let us consider only the second diagram of
Fig.~\ref{fig:mesons}, where the photon couples only to the lighter
vector meson $V$, and denote the initial and intermediate pseudoscalar
mesons as $M$ and $P$, respectively.  We restrict to this single
diagram because no positive measurements of $MM^* \gamma$ couplings
have yet appeared (recall that we used an upper bound for
$DD^*\gamma$), while $\Gamma (K^{*+} \to K^+ \gamma)$ and
$\Gamma(\rho^+ \to \pi^+ \gamma)$ are known.  This simple calculation
yields
\begin{eqnarray}
\Gamma (M \to V \gamma) & = & \frac 3 2 G_F^2 \left| V_M V_P \right|^2
f_M^2 f_P^2 \, B^2 \, \Gamma_{V \to P \gamma} \left[ \frac{{\cal C}
(M^2-m_P^2)}{{\cal C} (0)} \right]^2 \nonumber \\ & & \times \left(
\frac{M^2}{M^2-m_P^2} \right)^2 \left( \frac{M^2-m_V^2}{m_V^2-m_P^2}
\right)^3 \left( \frac{m_V}{M} \right)^3 , \label{rate}
\end{eqnarray}
where $B$ is the relevant bag parameter, the width $\Gamma_{V \to P
\gamma} \propto \alpha_{\rm EM}$, and the off-shell extrapolation of the
electromagnetic form factor, labeled ${\cal C}$, is explicitly
indicated.  Values for branching ratios for the six modes, along with
the photon energies, are listed in Table~\ref{ratenum}.

\begin{table}[t!]

\caption{\label{ratenum}Estimates of branching ratios for weak
annihilation decays using Eq.~(\ref{rate}).  Also included are
energies of the monochromatic photon.}

\vskip 2ex

\begin{center}
\setlength{\tabcolsep}{9pt}
\renewcommand{\arraystretch}{1.2}

\begin{tabular}{lccccc}
\tableline

Decay mode && BR (est.) && Photon Energy (GeV) & \\
\hline

$B^+   \to D^{*+}_s \gamma$ && $1 \times 10^{-7}$ && 2.22 & \\

$B^+   \to D^{*+}   \gamma$ && $7 \times 10^{-9}$ && 2.26 & \\

$B_c^+ \to K^{*+}   \gamma$ && $3 \times 10^{-6}$ && 3.14 & \\

$B_c^+ \to \rho^+   \gamma$ && $3 \times 10^{-5}$ && 3.15 & \\

$D^+   \to K^{*+}   \gamma$ && $6 \times 10^{-7}$ && 0.72 & \\

$D_s^+  \to \rho^+  \gamma$ && $8 \times 10^{-5}$ && 0.83 & \\
\hline

\end{tabular}
\end{center}
\end{table}

We see that the Cabibbo-unsuppressed decay $D_s^+ \to \rho^+ \gamma$
has a rate already large enough that it might already have been
produced at Fermilab or CLEO.  Certainly it will be produced copiously
at BABAR and BELLE, where also the rarer $B^+$ modes may be observed
in smaller but still significant numbers.  The $B_c$ channels must
necessarily wait for hadron machines such as LHC or BTeV.

One may also consider information contained in the helicity of the
photon.  For an arbitrary $P(0^-) \to V(1^-) \gamma$ decay, the
generic amplitude is
\begin{equation}
{\cal M} = \epsilon^{* \, (V)}_\mu \epsilon^{* \, (\gamma)}_\nu \left[
i {\cal A}_{PC} \, \epsilon^{\mu \nu \rho \sigma} p^{(V)}_\rho
p^{(P)}_\sigma + {\cal A}_{PV} \left( p^{(P) \, \mu} p^{(P) \, \nu} -
g^{\mu \nu} \, p^{(\gamma)} \! \cdot \! p^{(P)} \right) \right] ,
\end{equation}
where $PC$, $PV$ distinguish parity conserving and violating
amplitudes, respectively.  Then the total rate is
\begin{eqnarray}
\Gamma & = & \frac{1}{8\pi} | {\bf p} |^3 \left( \left| {\cal A}_{PC} +
{\cal A}_{PV} \right|^2 + \left| {\cal A}_{PC} - {\cal A}_{PV}
\right|^2 \right) \nonumber \\ 
& = & \frac{1}{4\pi} | {\bf p} |^3 \left( \left| {\cal A}_{PC}
\right|^2 + \left| {\cal A}_{PV} \right|^2 \right) , \label{PCPV}
\end{eqnarray}
where $|{\bf p}| = (m_P^2 - m_V^2)/2m_P$.  The first line of
Eq.~(\ref{PCPV}) is separated into contributions in which the two
vector particles are both right-handed (RR) and left-handed (LL) ,
respectively.  Indeed, the asymmetry is
\begin{equation}
\frac{\Gamma_{RR} - \Gamma_{LL}}{\Gamma} = \frac{2 {\rm Re} \, {\cal
A}_{PC} {\cal A}^*_{PV}}{\left| {\cal A}_{PC} \right|^2 + \left| {\cal
A}_{PV} \right|^2} \ .
\end{equation}
The relative weights of the two helicities may prove to be especially
interesting since the $V\!-\!A$ nature of weak interactions weights
the two photon helicities differently.  For example, in the case of
penguin $B^- \to K^{*-} \gamma$ and $\rho^- \gamma$ decays, the $L$
helicity has been found~\cite{GrossP} to be enhanced compared to $R$.
This enhancement persists~\cite{GrinP} even when long-distance
corrections (including contributions from $A$ diagrams) are included.
Certainly, a measured enhancement of the disfavored helicity would be
a signal of new physics.  Studies of the role of photon helicities are
also underway~\cite{CLP} in the pure $A$ decays described here.

The radiative weak annihilation decays occupy a unique position in
heavy flavor physics, in that they are completely flavor self-tagged
and kinematically trivial.  Their experimental observation is imminent
and promises another handle on the CKM matrix.  Once the most common
mode $D^{*+}_s \to \rho^+ \gamma$ is observed, its measured branching
ratio may be used to study the other decays.  Alternately, lattice
simulations may be used to probe the generalized bag parameters, one
may relate nonleptonic $A$ processes to semileptonic radiative modes
such as $B \to \gamma \ell \nu$~\cite{GrinP,KPY}, or one may consider
alternate new physics contributions.  On both the theoretical and
experimental fronts, many opportunities for advances exist.

\Acknowledgments
I thank the University of Maryland Theoretical Quarks, Hadrons, and
Nuclei group for their hospitality.


\begin{thebibliography}{99}


\bibitem{GHLR}
M.~Gronau, O.F.~Hernandez, D.~London, and J.L.~Rosner,
{\sl Phys.\ Rev.}  {\bf D50} (1994) 4529.

\bibitem{RFL}
R.F.~Lebed,
{\sl Phys.\ Rev.} {\bf D61} (2000) 033004.

\bibitem{PDG}
D.E.~Groom {\it et al.} (Particle Data Group),
{\sl Eur.\ Phys.\ J.} {\bf C15} (2000) 1.

\bibitem{AK}
P.~Asthana and A.N.~Kamal,
{\sl Phys.\ Rev.} {\bf D43} (1991) 278.

\bibitem{BGHP}
G.~Burdman, E.~Golowich, J.L.~Hewett, and S.~Pakvasa,
{\sl Phys.\ Rev.} {\bf D52} (1995) 6383.

\bibitem{KSW}
A.~Khodjamirian, G.~Stoll, and D.~Wyler,
{\sl Phys.\ Lett.} {\bf B358} (1995) 129.

\bibitem{Fajfer}
S.~Fajfer, S.~Prelov\v{s}ek, and P.~Singer,
{\sl Eur.\ Phys.\ J.} {\bf C6} (1999) 471;
S.~Fajfer and P.~Singer,
{\sl Phys.\ Rev.} {\bf D56} (1997) 4302;
B.~Bajc, S.~Fajfer, and R.J.~Oakes,
{\it ibid.} {\bf 51} (1995) 2230.

\bibitem{GL}
B.~Grinstein and R.F.~Lebed,
{\sl Phys.\ Rev.} {\bf D60} (1999) 031302(R).

\bibitem{AS}
T.M.~Aliev and M.~Savci,
{\sl J. Phys.} {\bf G24} (1998) 2223.

\bibitem{EGN1}
D.H.~Evans, B.~Grinstein, and D.R.~Nolte,
{\sl Phys.\ Rev.} {\bf D60} (1999) 057301.

\bibitem{EGN2}
D.H.~Evans, B.~Grinstein, and D.R.~Nolte,
{\sl Phys.\ Rev.\ Lett.} {\bf 83} (1999) 4947.

\bibitem{EGN3}
D.H.~Evans, B.~Grinstein, and D.R.~Nolte,
{\sl Nucl.\ Phys.} {\bf B577} (2000) 240.

\bibitem{GrossP}
Y.~Grossman and D.~Pirjol,
{\sl JHEP} {\bf 0006} (2000) 029.

\bibitem{GrinP}
B.~Grinstein and D.~Pirjol,
{\sl Phys.\ Rev.} {\bf D62} (2000) 093002.

\bibitem{CLP}
C.E.~Carlson, R.F.~Lebed, and A.~Petrov, in preparation.

\bibitem{KPY}
G.P.~Korchemsky, D.~Pirjol, and T.-M.~Yan,
{\sl Phys.\ Rev.} {\bf D61} (2000) 114510.

\end{thebibliography}
\end{document}